\documentclass[sigconf]{acmart}

\usepackage{framed}
\usepackage{xcolor}
\usepackage{multirow}
\definecolor{framecolor}{rgb}{0.8,0.2,0.2} 
 {\endMakeFramed}

\AtBeginDocument{%
  }

\setcopyright{acmlicensed}
\copyrightyear{2018}
\acmYear{2018}
\acmDOI{XXXXXXX.XXXXXXX}

\acmConference[Conference acronym 'XX]{Make sure to enter the correct
  conference title from your rights confirmation emai}{June 03--05,
  2018}{Woodstock, NY}
\acmISBN{978-1-4503-XXXX-X/18/06}

\begin{document}

\title[Ashes or Breath]{\textit{Ashes or Breath}: Exploring Moral Dilemmas of Life and Cultural Legacy through Mixed Reality Gaming}

\author{Black Sun}
\affiliation{%
 \institution{Department of Computer Science \\ Aarhus University}
 \city{Aarhus}
 \country{Denmark}
 }
\email{blackthompson770@gmail.com}

\author{Ge Kacy Fu}
\affiliation{%
 \institution{Department of Computer Science\\ Aarhus University}
 \city{Aarhus}
 \country{Denmark}
 }
\email{kacyfu01@gmail.com}

\author{Shichao Guo}
\affiliation{%
 \institution{Department of Computer Science\\ Aarhus University}
 \city{Aarhus}
 \country{Denmark}
 }
 \email{g18603914990@foxmail.com}



\renewcommand{\shortauthors}{B. Sun et al.}


\begin{abstract}
Traditional approaches to teaching moral dilemmas often rely on abstract, disembodied scenarios that limit emotional engagement and reflective depth. To address this gap, we developed \textit{Ashes or Breath}, a Mixed Reality game delivered via head-mounted displays(MR-HMDs). This places players in an ethical crisis: they must save a living cat or a priceless cultural artifact during a museum fire. Designed through an iterative, values-centered process, the experience leverages embodied interaction and spatial immersion to heighten emotional stakes and provoke ethical reflection. Players face irreversible, emotionally charged choices followed by narrative consequences in a reflective room, exploring diverse perspectives and societal implications. Preliminary evaluations suggest that embedding moral dilemmas into everyday environments via MR-HMDs intensifies empathy, deepens introspection, and encourages users to reconsider their moral assumptions. This work contributes to ethics-based experiential learning in HCI, positioning augmented reality not merely as a medium of interaction but as a stage for ethical encounter.

\end{abstract}




\maketitle

\section{Introduction}

Moral reasoning and ethical reflection are essential competencies in civic education and cultural engagement \cite{kohlberg1981philosophy, dewey2024democracy}. Traditionally, they are taught through abstract dilemmas such as the Trolley Problem, which, while intellectually stimulating, often lack emotional or embodied engagement \cite{greene2014moral}. This absence of situated, affective experience limits the development of empathy, contextual sensitivity, and moral imagination. To address this gap, we introduce \textit{Ashes or Breath}, a Mixed Reality (MR) experience delivered via head-mounted displays (MR-HMDs). During a simulated museum fire, players must choose between rescuing a living cat or a priceless cultural artifact. Unlike traditional exercises, this scenario requires immediate, embodied action in a spatially immersive environment, transforming ethical reasoning into felt responsibility. \textit{Ashes or Breath} emphasizes ambiguity, emotional resonance, and cultural reflection. Gameplay unfolds in two stages: an immersive museum scene where players face the moral decision, followed by the ``Rewind Room,'' where memory fragments reveal societal and personal consequences. This design encourages players to reconsider their stance and engage in critical reflection.  

Our contributions are (1) a case study of a culturally informed MR game that provokes ethical decision-making; and (2) empirical insights into reflection and emotional engagement with moral dilemmas in MR contexts. This work positions MR as an interaction medium and an ethical stage, where dilemmas become lived tensions that demand responsibility rather than resolution.

\section{Related Work}

\subsection{Moral Dilemma Games and Embodied Ethical Decision-Making}

Digital games have long explored ethical ambiguity, providing controlled environments to confront conflicting values such as empathy, justice, and self-interest \cite{sicart2011ethics, belman2010designing}. Classic titles like \textit{Papers, Please}\footnote{\url{https://store.steampowered.com/app/239030/Papers_Please}} and \textit{This War of Mine}\footnote{\url{https://store.steampowered.com/app/282070/This_War_of_Mine}} immerse players in high-stakes decisions without clear right or wrong answers. HCI research shows that emotional salience and embodied interaction can further heighten such dilemmas’ educational and reflective potential \cite{daneels2021eudaimonic, schrier2017designing, holl2020moral, zhang2024exploring}. Work on realism and ecological validity demonstrates how urgency and presence alter moral trade-offs. \citeauthor{jang2024whom} introduce time-critical VR dilemmas that reveal the effect of cascading decisions under pressure \cite{jang2024whom}. \citeauthor{niforatos2020would}~\cite{niforatos2020would} show that enacting classic scenarios in VR amplifies utilitarian responses and reveals cultural and gender variability. At a meta-level, \citeauthor{nunes2022scoping}~\cite{nunes2022scoping} review SIGCHI research and highlight growing demand for embodied, reflective ethical inquiry, often through participatory and value-sensitive design. Together, these works suggest that immersive technologies do more than present ethical scenarios; they reshape how users interpret and act on them, fostering accountability and immediacy. \textit{Ashes or Breath} builds on this literature by embedding MR into moral dilemma design, complicating choices emotionally and offering structured reflection. Its dual structure, immersive dilemma followed by reflective narrative, provides a novel template for value-driven MR experiences.  

\subsection{Cultural Heritage and Interactive Media in HCI}

The convergence of intangible cultural heritage (ICH) and interactive media has become a key HCI concern, as emerging technologies are used to preserve, revitalize, and reinterpret cultural practices. Immersive systems in particular enable embodied and affective engagement with ICH. For example, Li et al.’s \textit{DianTea} system supports youth learning of traditional tea rituals through VR and game-based pedagogy~\cite{li2023diantea}, \citeauthor{wang2025facilitating}~\cite{wang2025facilitating} designed a VR system for Chinese flower arrangement that sustains practice through reflection and multisensory immersion. Other approaches emphasize co-creation: \citeauthor{liu2023digital}~\cite{liu2023digital} used digital making to reimagine folk handicrafts, and Wang et al. \cite{wang2024critical} applied Critical Heritage Studies to Douyin, showing how short-video platforms both democratize and complicate heritage discourse across identity and knowledge domains. \textit{Ashes or Breath} extends this line of inquiry by framing cultural heritage as content to be preserved and a contested moral value. By forcing a choice between saving a living being or a priceless artifact, it foregrounds the tension between empathy and legacy, aligning with speculative and critical design approaches~\cite{dunne2008hertzian}.

\section{GAME DESIGN AND IMPLEMENTATION}

\subsection{Design Purpose}

The goal of \textit{Ashes or Breath} is to explore how MR can support embodied ethical reflection through situated moral dilemmas. Rather than presenting ethics as abstract reasoning detached from lived experience, the game embeds moral decision-making directly into the player’s physical environment, demanding action shaped by spatial proximity, emotional immediacy, and personal accountability. This immersive framing enables moral reflection to unfold not only as a cognitive process but as an affective and embodied one, grounded in the tensions of real-world interaction. Players must navigate a choice juxtaposing the preservation of cultural heritage against protecting a cat's life; no option is morally unambiguous or free of consequence.


By immersing players in morally fraught scenarios, \textit{Ashes or Breath} aims to advance a range of educational outcomes: fostering moral reasoning through value conflict and trade-offs \cite{hyry2022moral}; cultivating empathy and the capacity to understand diverse perspectives; encouraging critical thinking through engagement with uncertain, high-stakes decisions \cite{van1995car}; and prompting self-reflection on personal values and ethical commitments \cite{makransky2018structural, belman2010designing, taylor2019ethical}. In doing so, the research positions MR not merely as a delivery mechanism for ethical content but as a transformative pedagogical medium capable of deepening moral awareness through experiential learning \cite{niforatos2020would}.

\begin{figure*}[htbp]
    \centering
    \includegraphics[width=1\textwidth]{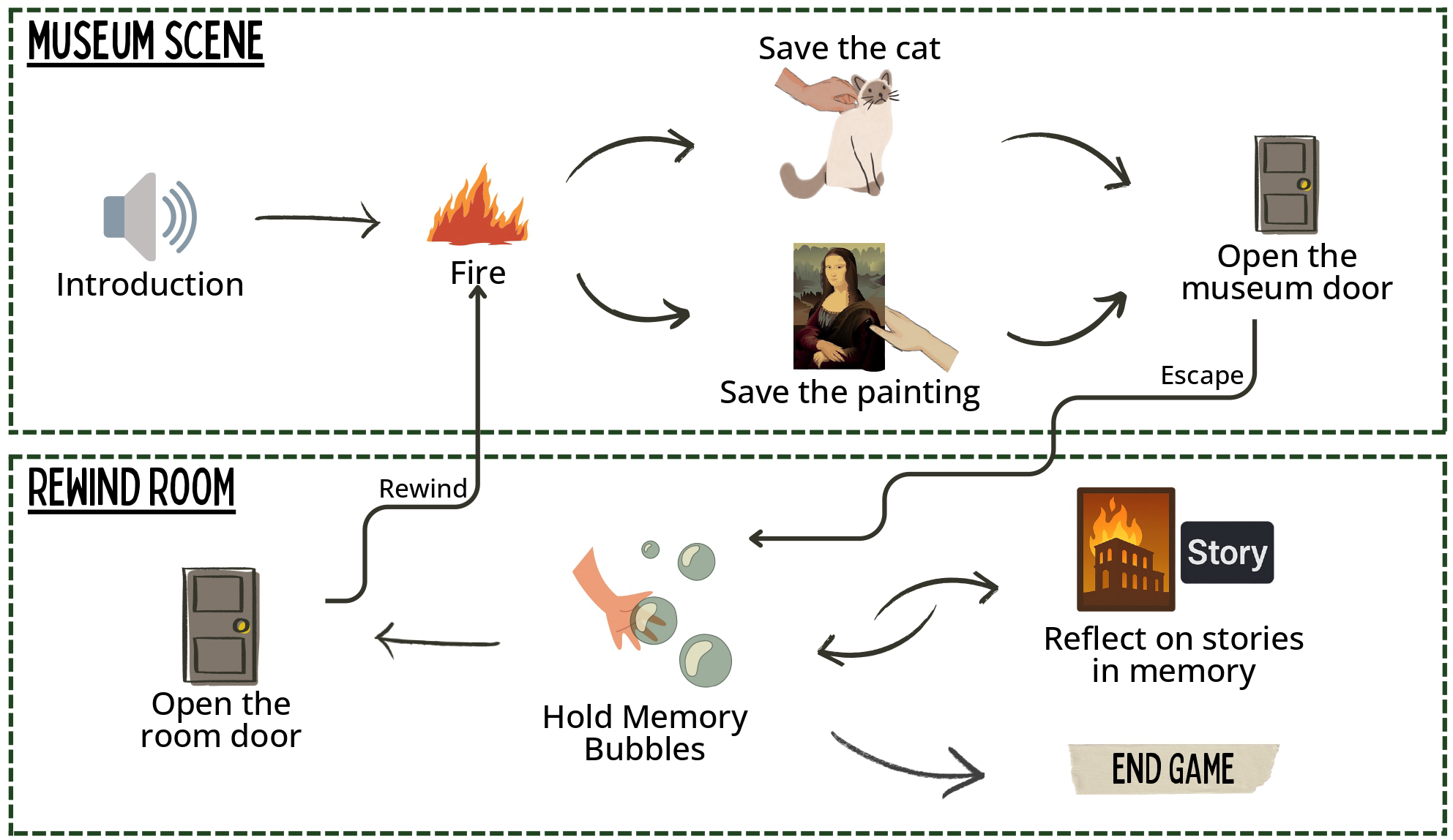}
    \caption{Ashes or Breath Game Flow.}
    \label{fig:gameflow}
\end{figure*}

\subsection{Game Design}

\subsubsection{Museum Scene}

\begin{figure*}[htbp]
    \centering
    \includegraphics[width=1\textwidth]{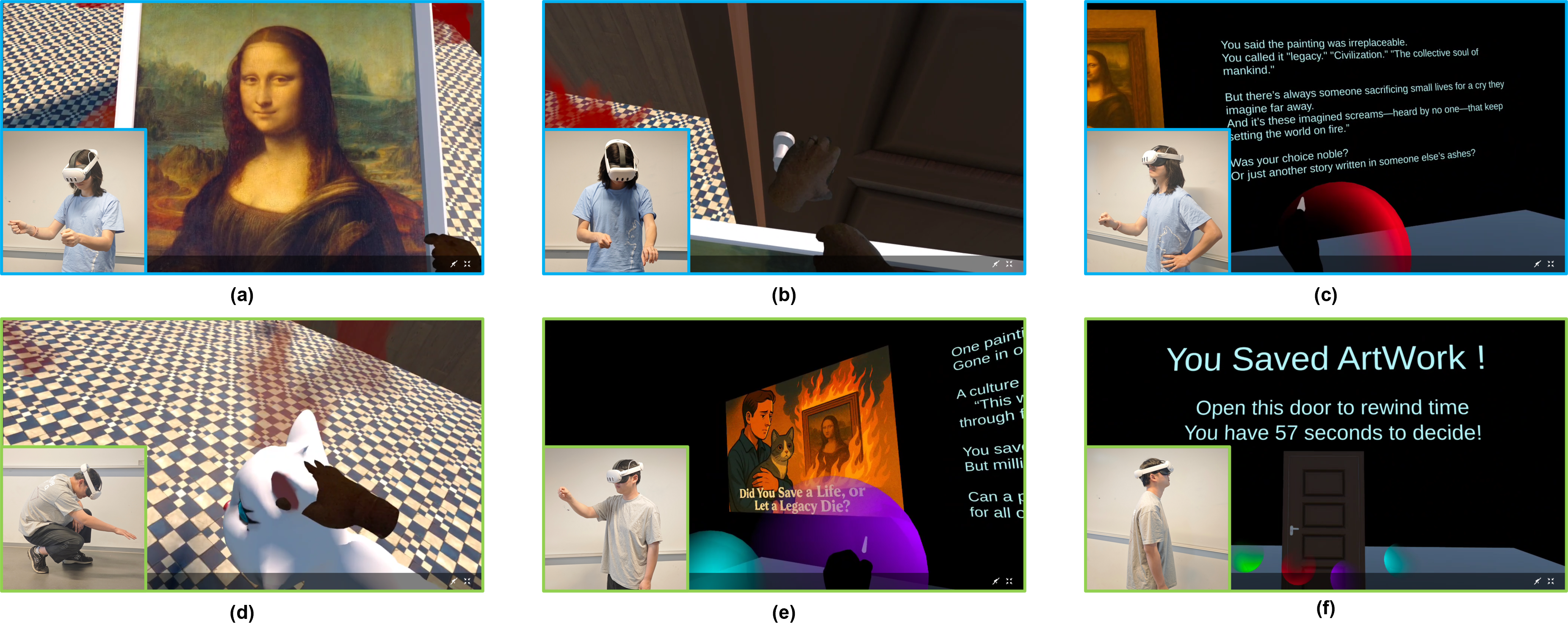}
    \caption{Illustrative gameplay moments in \textit{Ashes or Breath}. (a–c) The player chooses to save the painting and exits the museum through the door, entering the \textit{Rewind Room}. (d–e) The player instead chooses to save the cat and proceeds into the \textit{Rewind Room}. (f) Within the \textit{Rewind Room}, the player can interact with the door to return to the museum for another decision.}
    \label{fig:your_label}
\end{figure*}

As shown in Fig.~\ref{fig:gameflow}, \textit{Ashes or Breath} unfolds in two phases: an interactive decision space and a reflective narrative space. In the first phase, players enter a digitally augmented museum via a Mixed Reality headset. Two focal elements are presented: a culturally significant painting (we use the famous painting \textit{Mona Lisa} by Leonardo da Vinci) and a responsive cat. At the beginning of the experience, an audio introduction about the painting plays within the museum space. During this time, the virtual cat idly roams the museum floor, either idling or walking. Players can interact with the painting and the cat while exploring the environment. Suddenly, the calm is disrupted by a fire. Players must make a choice, saving either the cat or the painting, thus committing to one ethical priority. This moral choice is designed to be emotionally challenging and temporally constrained, heightening the stakes of the experience. The system provides no score, hint, or judgment, emphasizing internal justification over external reward. Once players have grabbed either the painting or the cat, they can run toward the museum door, escape to transition into the \textit{Rewind Room} scene (See Fig.~\ref{fig:your_label}(a-e)).

\subsubsection{Rewind Room}

Following the decision, the game transitions into the second phase: the \textit{Rewind Room}. To encourage ethical reflection beyond the moment of decision, the \textit{Rewind Room} presents four distinct “memory bubbles,” each visualizing a different dimension of the consequences of the player's choice using figures and thought-provoking text(See Fig.~\ref{fig:your_label}(c,e)). These bubbles are designed to provoke critical thinking, perspective-taking, and emotional engagement by revealing the complex ripple effects of either action or inaction: (1) \textbf{Immediate Aftermath:} Presents short-term consequences, such as media headlines or public reactions, that frame the player's decision within broader moral discourse. (2) \textbf{Values Reflection:} Prompts players to reflect on their choices' internal logic and moral assumptions, often through contrasting ethical perspectives. (3) \textbf{Personal Emotional Impact:} Depicts the lingering psychological effects over time, including emotions such as pride, affirmation, guilt, doubt, or regret. (4) \textbf{Broader Societal Consequences:} Highlights long-term cultural and social responses, including shifts in collective memory, public discourse, and debates around the value of heritage versus life.

Notably, these bubbles emphasize the path not taken; if the player rescues the cat, one memory may lament the irreversible loss of a cultural masterpiece; if they choose the painting, a haunting dream might evoke the cat's final moments. This asymmetrical narrative strategy is designed to provoke critical thinking, ethical reflection, empathy, and perspective-taking by rendering the invisible cost of every decision both visible and emotionally resonant. Rather than reinforcing moral binaries, the \textit{Rewind Room} foregrounds complexity, ambiguity, and the inherent trade-offs that define ethical life. If players are satisfied with their decision, the experience ends at this point. However, if they feel regret or wish to reconsider their choice, they can open the door to transition back into the post-fire Museum Scene(See Fig.~\ref{fig:your_label}(f)), where they can make a different decision. This looping structure can repeat indefinitely, allowing players to re-enter the dilemma and explore its consequences from multiple ethical perspectives. The experience concludes only when the player has reflected sufficiently and chooses to end the game.


\subsection{Game Implementation}

We developed \textit{Ashes or Breath} in Unity3D with Meta XR Tools’ Building Blocks (e.g., Hand Tracking, Real Hands, Scene Mesh) to enable spatial interaction and MR functionality. 3D models were sourced from the Unity Asset Store\footnote{\url{https://assetstore.unity.com}} and Sketchfab\footnote{\url{https://sketchfab.com}}, while textures came from Poly Haven\footnote{\url{https://polyhaven.com/textures}}. The \textit{Mona Lisa} and consequence images in the \textit{Rewind Room} were generated using GPT-4o\footnote{\url{https://chatgpt.com}}. The game runs on Meta Quest 3 and Quest 3s headsets. In the Museum Scene, the headset scans the physical environment and anchors the \textit{Mona Lisa} to a vertical surface using wall-based scene understanding. Interactions rely on embodied gestures and spatial navigation: players grasp the cat or painting with their real hands, and physically push the door handle to transition between spaces.

\begin{table*}[ht]
\centering
\caption{Participants’ responses on playability, emotion, embodiment, and reflection in \textit{Ashes or Breath} (1 = Not at all, 5 = Extremely)}
\label{tab:user_experience}
\begin{tabular}{p{3cm}p{9cm}cc}
\toprule
\textbf{Aspect} & \textbf{Question} & \textbf{Mean} & \textbf{SD} \\
\midrule

Playability 
& Q1: I would consider the game worth playing. 
& 4.5 & 1.35 \\
\cmidrule(lr){1-4}

Emotional Intensity 
& Q2: I felt emotionally immersed when the museum caught fire. 
& 4.5 & 1.18 \\
& Q3: I was emotionally affected by seeing the cat "dead". 
& 3.7 & 1.16 \\
& Q4: I was emotionally affected by seeing the painting "burnt". 
& 3.5 & 1.08 \\
\cmidrule(lr){1-4}

Embodied Interaction 
& Q5: Interacting with virtual objects using my real hand felt immersive. 
& 4.1 & 0.74 \\
& Q6: Freely walking and interacting in the game felt immersive. 
& 4.3 & 0.75 \\
& Q7: Seeing the cat walk on the floor felt immersive. 
& 3.9 & 0.74 \\
\cmidrule(lr){1-4}

Empathy and Reflection 
& Q8: The memory bubbles made me rethink my initial decision. 
& 3.4 & 1.58 \\
& Q9: The game helped me understand the “other side” of the dilemma. 
& 4.1 & 0.88 \\
& Q10: The game helped me grow in moral sensitivity and empathy. 
& 3.7 & 1.16 \\
\bottomrule
\end{tabular}
\end{table*}

\section{Evaluation and Findings}

\subsection{Study Setup}

We conducted a user study to evaluate emotional and ethical engagement in the experience. We focused on how players responded to the core moral dilemma rather than mechanical performance or usability alone. The study included 12 participants (5 female, 7 male, P1–P12), aged 21–35, with backgrounds in digital design, computer science, and the general public. Seven participants had never experienced a moral dilemma game prior to the study. Regarding MR familiarity, one participant had no prior experience with MR-HMDs, three had little experience, and six were very familiar with MR-HMDs. Each session lasted approximately 10–15 minutes, including gameplay and post-play reflection. Participants experienced the complete MR Game \textit{Ashes or Breath} using Meta Quest 3 with hand tracking and then completed a post-play survey. The survey included both \textit{Likert-scale} items (1-5) and open-ended questions designed to assess: (1) Emotional intensity at key moments (e.g., onset of fire); (2) Strength of embodied presence and gesture impact; (3) Perceived consequences and retrospective reflection.

We analyzed open responses using an iterative thematic analysis \cite{braun2006using}, identifying seven sub-themes and three main themes related to presence, ethical reasoning, and emotional resonance. Two researchers independently coded the data and aligned their categories to ensure consistency.

\subsection{Quantitative Results}

Five participants chose to save the cat, and five decided to save the painting. All participants made at least two decisions (i.e., entered the Rewind Room twice). Two participants ultimately changed their initial choice, while seven others reported better understanding the perspective behind the option they did not select. We collected survey responses across four dimensions: playability, emotional intensity, embodied interaction, and moral reflection. As summarized in Table~\ref{tab:user_experience}, participants rated the game as emotionally immersive and physically engaging, particularly during the fire event and through hand-tracked interaction. Reflection-related items received slightly lower but positive responses, indicating that the Rewind Room successfully prompted reconsideration and perspective-taking, though it needed more improvement.

\subsection{Qualitative Results}

\subsubsection{Emotional Response and Moral Engagement}

The feedback indicated that the game's emotional arc successfully guided players from calm observation to ethically loaded tension. In the Museum Scene, participants noted the immersive serenity of the MR environment, describing it as immersive and engaging. Most participants reported a clear emotional shift when the fire broke out, marking the transition from observation to ethical urgency. P3 described the moment as \textit{"an emotional punch—suddenly, it wasn’t a simulation anymore."} P1 and P6 both hesitated before acting, with P6 stating, \textit{"I just stood there thinking and feeling the pressure."} Participants frequently expressed feelings of guilt or regret, regardless of their choice. P2, who saved the cat, said, \textit{"I wanted to save both, but the cat looked at me. I couldn’t leave it."} Meanwhile, P7, who chose the painting, noted, \textit{"I convinced myself Mona Lisa mattered more. But I’m still not sure I believe that."} These emotionally charged responses suggest that the game successfully elicited moral tension without prescribing a correct path.

\subsubsection{Perceived Value and Ethical Reasoning}

The game was interpreted as a vehicle for ethical introspection, with players drawing connections to broader themes of cultural preservation, utilitarian reasoning, and empathy. Participants reflected on their choices using language grounded in ethics, empathy, and value trade-offs. P4 described the decision as a “test of instinct versus legacy,” while P9 noted, \textit{"It felt like a stand-in for bigger moral questions (like climate, or war) where you can't save everything."} The Rewind Room was consistently described as reshaping player understanding. P1, P5, and P10 each mentioned that narrative feedback prompted reevaluation of their original choice. P5 said, \textit{"The second time, I saw the grief bubble and realized I wasn’t being honest the first time."} P3, P7, and P9 indicated that the looping structure made them more aware of their internal justifications. P3 wrote, \textit{"I thought I was logical. Then I realized I was avoiding something."} Themes around the \textit{"symbolic weight of art"} (P8), the \textit{"duty to living beings"} (P2, P6), and the \textit{"emotional cost of irreversible decisions"} (P4) emerged across responses, indicating that the experience activated deep value-based reasoning.

\subsubsection{Interaction and Playability}

Technically, most participants found the AR scene intuitive and engaging. P3 and P6 described the hand tracking as “natural,” while P8 called it “immersive and believable.” However, participants P1, P5, and P7 noted confusion during the transition to the fire event, suggesting that clearer affordances or visual cues could support the pacing. Narratively, P10 asked about the cat's background or the artwork, expressing curiosity about the story context. P4 asked, \textit{"Is the cat someone's pet? I wanted to know more before deciding."} This feedback points to the potential of embedding optional lore or interactive narrative fragments to support richer emotional grounding.

\section{DISCUSSION}

Our findings highlight how Mixed Reality can be a powerful medium for engaging users in situated, embodied ethical reflection. In this section, we synthesize key themes across four interrelated areas: ambiguity as moral design material, embodied ethical tension in MR, cultural value conflict, and replayability. These insights point toward actionable design implications for future values-centred and immersive HCI work.

\subsection{Situated Moral Ambiguity and Embodied Ethics}

A central observation is that players did not seek to “solve” the moral dilemma, but to inhabit and navigate its ambiguity. In contrast to games that assign apparent success or failure metrics, our design emphasised unresolved value conflict. This ambiguity prompted players to reflect on their ethical assumptions, supporting the view that uncertainty can enhance interpretive engagement \cite{gaver2003ambiguity, bentvelzen2022revisiting}. This ethical tension was further intensified through embodied MR interaction. Hand-tracked gestures and spatial immersion elevated the stakes of the decision, transforming abstract judgment into immediate physical action. This aligns with prior work on embodied cognition and affective presence in mixed-reality systems \cite{marin2020emotion}. Participants described these moments as emotionally “visceral,” reinforcing the potential of MR to evoke heightened moral salience when decisions are enacted within users’ lived environments.

\subsection{Culture as a Site of Ethical Conflict}

Beyond representing cultural heritage, our design framed it as a moral stake. The decision to sacrifice a living being or a cultural artifact positioned heritage not as neutral content but as an active site of value contestation. This approach diverges from heritage games that focus on preservation, and instead surfaces the affective tensions between legacy and empathy. This framing echoes calls in speculative and feminist design to interrogate whose values are embedded in interactive systems critically \cite{bardzell2010feminist}. Rather than resolving these tensions, our game maintains instability, encouraging players to confront conflicting priorities without a clear resolution. Culture here is not static, but emotionally and ethically charged.




\subsection{Limitations and Future Work}

This study presents several limitations that inform directions for future exploration. First, our evaluation was based on a limited number of participants in peer and showcase environments, which restricts the generalizability of findings. Broader, cross-cultural studies are needed to assess how users from diverse backgrounds engage with embodied moral dilemmas. Second, while the looping structure encouraged replay and reflection, it may also reduce the perceived weight of the original decision. Enhancing consequence persistence or integrating delayed feedback could help balance reflection with ethical tension.

\section{CONCLUSION}

We presented \textit{Ashes or Breath}, a Mixed Reality (MR) game
using MR-HMDs that explore ethical decision-making at the intersection of cultural heritage and human empathy. By placing users in a lifelike dilemma—saving a cat or a cultural artifact—the experience blends embodied interaction with reflective narrative to surface moral tension and introspection. Our findings suggest that ethical ambiguity, spatial presence, and irreversible action can foster deep engagement and emotional reflection. Rather than resolving moral questions, this work positions MR as a medium for staging them, inviting users to confront discomfort, uncertainty, and responsibility. This contribution highlights the potential of MR for ethics-based experiential learning in civic and cultural contexts. Future work will expand narrative diversity, explore longitudinal impact, and investigate applications in education and public engagement.

\bibliographystyle{ACM-Reference-Format}
\bibliography{reference}

\end{document}